\documentclass[fleqn,twoside]{article}
\usepackage{espcrc2}

\usepackage{graphicx}
\usepackage[figuresright]{rotating}

\newcommand{\AmS}{{\protect\the\textfont2
  A\kern-.1667em\lower.5ex\hbox{M}\kern-.125emS}}

\def\lesssim{\mathrel{\hbox{\rlap{\hbox{\lower4pt\hbox{$\sim$}}}\hbox{$<$}}}}
\def\gtrsim{\mathrel{\hbox{\rlap{\hbox{\lower4pt\hbox{$\sim$}}}\hbox{$>$}}}}
\newcommand{\ffrac}[2]{\left(\frac{#1}{#2} \right)}

\title{Antimatter Bounds by Anti-Asteroids annihilations on Planets and Sun}

\author{D.Fargion \address[Uniroma1]
        {Physics department, Universita' degli studi "La Sapienza", \\
        5, Piazzale Aldo Moro - I 00185 Roma, Italy}%
        \address[INFN]
        {INFN Roma, Istituto Nazionale di Fisica Nucleare, Italy}%
        and M.Khlopov        \address[KHLO]
        {Centre for Cosmoparticle physics "Cosmion", \\
        Miusskaya Pl. 4, 125047, Moscow, Russia}%
        \address[IPM]
        {Keldysh Institute of Applied Mathematics, \\
      Miusskaya Pl. 4, 125047, Moscow, Russia}%
}

\begin{document}

\begin{abstract}
 The existence of antimatter stars in the Galaxy as possible signature
 for inflationary models with non-homogeneous baryo-synthesis may leave the trace
 by antimatter cosmic rays as well as by their secondaries (anti-planets and
 anti-meteorites) diffused bodies in our galactic halo. The anti-meteorite flux may leave its
 explosive gamma signature by colliding on lunar soil as well as on
 terrestrial,jovian and solar atmospheres.
  However the propagation in galaxy and the consequent
 evaporation in galactic matter gas suppress the lightest ($m< 10^{-2} g$) anti-meteorites.
 Nevertheless heaviest anti-meteorites ( $m >$ $10^{-1}$ g up to $10^{6}$ g)
 are unable to be deflected or annihilate by the thin galactic gas surface annihilation;
 they might hit the Sun (or rarely Jupiter) leading to an explosive gamma event and a spectacular
 track with a bouncing and even a propelling annihilation on cromosphere and photosphere.
 Their anti-nuclei annihilation in pions and their final hard gammas showering
 may be observable as a "solar flare" at a rate nearly comparable to the observed ones.
 From their absence we may infer first bounds on antimatter-matter ratio near
 or below $10^{-9}$ limit applying  already recorded   data in gamma BATSE catalog.

\vspace{1pc}
\end{abstract}

\maketitle

\section{INTRODUCTION}

Severe constrains on the possibility of baryon symmetrical
Universe (see review in [1-3]), as well as the evident baryon
asymmetry of our cosmic neighborhood, related in the modern
cosmology to the process of baryosynthesis in the very early
Universe ([4], see e.g.[5] for review), do not exclude the
existence of relatively small amount of sufficiently large
regions of antimatter in the modern Universe, reflecting the
nontrivial physical processes, underlying inflation and
baryosynthesis. The original idea [6,3,5] to consider antimatter
in the baryon asymmetrical Universe as the tracer for the strong
nonhomogeneity of baryosynthesis finds support in recently
developed inflationary models with nonhomogeneous spontaneous
baryosynthesis [7]. Such models reproduce in quantitative way
both the possibility of diffused antiworld (regions of very low
density antiproton-positron plasma) [8] and the hypothesis on the
existence of antimatter stars in our Galaxy [9]. The both
possibilities satisfy the severe constrains on matter-antimatter
annihilation [1-3]. In particular, the antimatter globular
cluster as the possible form of antimatter in our Galaxy is
consistent with that constrains, since the substantial growth of
annihilation zone and depletion zone at matter-antimatter
boundary at redshift $z = 3$ was found in [1] for the case of
large domains in baryon symmetrical Universe. According to [1]
this result is not applicable to the case of small (about
$10^{-6}$) relative volume, occupied by antimatter in
baryon-asymmetrical Universe, when the size of antimatter
domains, surviving to the present time, is determined as in [8].
At the enhanced density of antibaryons in domain  it provides
formation of globular cluster of antimatter stars [9]. Moreover,
it was shown recently [10] that annihilation of antimatter, lost
by antimatter stars in the form of stellar wind, can reproduce
the observed galactic gamma background in the range tens-hundreds
MeV. Still any source of neutral pions can lead to the same
effect and the manifest signature for existence of antimatter
stars is the existence of antinuclear component of cosmic rays,
accessible to the future cosmic ray experimental searches, first
of all in AMS-II experiment. The other profound signature of
antimatter are the pieces of antimatter, coming in the form of
antimatter meteorites. We study the latter possibility in the
present paper and find it interesting tool to probe the origin of
matter, related with the creation of antimatter. With all the
 uncertainties and reservations, taken into account, the search
for antimatter meteorites can still provide the useful probe for
 the existence of macroscopic antimatter.
 \section{GAMMA FLASHES by
ANTIMETEORITE ANNIHILATIONS
  ON EARTH and MOON}
 The present flux of meteorites with the mass $M$ observed on the Earth is nearly
$10^{4}\ffrac{M}{10kg}^{-1}$ event a year. This power extend
 for a large range of mass values. It is very possible that most of
 this matter has a local "solar" origin. However simple argument
 on nearby stellar encounters and matter exchange imply that up to
 1\% of the meteorites may be of galactic (extra-solar) origin. Therefore up to
 nearly
 \begin{equation}
 \label{eq1}
 \frac{dN}{dt} = 10^6 \ffrac{M}{1g} ^{-1}
\end{equation}
of meteorites, hitting the Earth any year, can be of galactic
(extra-solar) nature. If the corresponding antimeteorites rate
follows the same power law, at any given suppressed ratio, $r$,
\begin{displaymath}
 r= \ffrac {N_{a}}{N_{m}}
\end{displaymath}
where $N_{a(m)}$ the total amount of antibaryons (baryons) in the
Galaxy, (let say a part over a million or a billion or below)
  its signal will be anyway power-full enough to be
  (in most cases) observable.
Indeed the amount of energy released during the annihilation
 follows common special relativity; for any light (milligram unit)
 anti-meteorites mass $M$ the energy ejected is :
\begin{equation}
 \label{eq2}
 E= 10^{18} \ffrac{M}{1mg} erg
\end{equation}
 its corresponding "galactic" event rate, following eq.(1) is
\begin{equation}
 \label{eq3}
 \frac{dN}{dt} = 10^9 r \ffrac{M}{1mg} ^{-1} year^{-1}
\end{equation}
The event of the anti-meteorite annihilation on the Earth
atmosphere will give life to unexpected upward gamma shower that
will mimic mini nuclear atomic test or extreme upward Gamma
Shower. Even for a large suppression ratio $r= 10^{-9}$ this event
rate derived from expression above  (one a year) should not
escape the accurate BATSE ten-year monitoring. Actually the
atmosphere area below BATSE detection is nearly 1\% of all Earth
leading to a total probability rate of 0.1 in ten years. However
the corresponding secondaries gamma flux by consequent nuclei
annihilation showering into charged and neutral pion and their
decays and degradation in atmosphere should lead to a huge gamma
fluence $F$ observable in a near orbit satellite as Beppo-Sax or
GRO Batse:
\begin{displaymath}
 F \simeq 10~ erg/cm^2 (M/1mg)
\end{displaymath}
\begin{displaymath}
 Flux = 100 ~ erg\cdot sec^{-1} cm^{-2}
\end{displaymath}
This latter flux is derived assuming a characteristic galactic
velocity v= 300 km/sec for the incoming anti-meteorite and a
terrestrial atmosphere of nearly 30 km height. Such a signal is
nearly 10 order of magnitude above the sensitive Batse detection
threshold. Smaller scale upward gamma flash are indeed known and
they are called "Terrestrial Gamma Flashes". They are
corresponding to just $10^{8}$ or $10^9$ erg of isotropic fluence
energy (or even much less energy if originated by beamed upward
$\tau$ airshowers at $10^{15} eV$ up to nearly horizontal ones at
$10^{19} eV$ \cite{Fargion2001}) released at millisecond up to
ten of second timescales. Therefore such milligram anti-meteorite
bang will be already loudly recorded on data, if they were taking
place. Of course so high large event fluence would not escape
also other less sensitive astrophysical or military detectors.
Therefore it seem that milligram antimatter meteorite rain should
be totally excluded at very low level ($ r \leq 10^{-9}$). Even
more dramatic and sharp gamma signature should come by their fast
Moon annihilation (because of the absence of atmosphere), but at
a less (Moon surface over Earth one) rate. Lunar anti-meteorite
annihilation in characteristic nano-second signature, would make
very strong signals at lunar orbiting gamma detectors. They
provide a complementary tool to exclude very light (micro-gram)
antimeteorite rains at the same severe bound ($ r \leq 10^{-9}$).

\section{LIGHT ANTIMETEORITE EVAPORATION  CROSSING THE GALAXY}

However these results may be alleviated keeping in mind that
antimeteorites can be annihilated or "evaporated" during their
propagation in galactic gas. Indeed the
column density of
atoms (protons) crossed assuming $n_{disk}= 1 \cdot cm^{-3}$ and
a galactic disk height of $h = 100$ pc and a total number of
crossing 100 is: $N= 3 \cdot 10^{22}$ $cm^{-2}$. Each crossed
matter atom annihilates on the surface of the rigid body of
anti-meteorite. Putting the total mass of the crossed matter gas
equal to the mass of spherical homogeneous antimeteorite of radius
$r$ and internal density $\rho$,
\begin{displaymath}
 \pi r^{2} N m_{H} = \frac{4 \pi}{3} \rho  r^{3}
\end{displaymath}
one obtains that the antimeteorite can not escape complete
annihilation, if its radius is smaller, than
\begin{displaymath}
 r_{an} =  \frac{3}{4} \cdot \frac {N m_{H}}{\rho}
\end{displaymath}
 and the corresponding meteorite mass, given by
\begin{displaymath}
M_{an} = \frac{9}{16} \pi \cdot \frac {(N m_{H})^{3}}{\rho^{2}}
\end{displaymath}
is  (assuming water density) about $2.2 \cdot 10^{-4}$ g. The
actual value of minimal mass of the antimeteorite, surviving
annihilation, may be a few orders of magnitude larger. If we take
into account the strong (cubic) dependence of $M_{an}$ on $N$, we
find important the increase of $N$ due to effects of annihilation
with the gas above the disc.
 The mass of antimeteorite, which is completely destroyed by annihilation, can
be even larger, if we take  into account its atomic composition.
To destroy the antimeteorite, which consists of anti-atoms with
atomic number $A$, it is not necessary to annihilate all the
anti-nucleons in all its antinuclei, since even the result of one
proton anti-nucleus annihilation not only destroys the
anti-nucleus, but also causes the successive destructive effects
by its fragments. We discuss the effects of energy and momentum
transfer due to such processes in the next section, and only
estimate here the increase in the minimal mass of anti-meteorite,
surviving after annihilation. Putting the total number of matter
gas atoms, annihilating on the surface of anti-meteorite, equal to
the total number of anti-atoms with atomic number $A$ in
antimeteorite, we obtain instead of $M_{an}$ the magnitude
\begin{equation}
M_{surv} = \frac{9}{16} \pi \cdot \frac {(A N m_{H})^{3}}{\rho^{2}}
\end{equation}
which is the factor of $A^{3}$ larger, than  $M_{an}$.
 This imply that milligram (and even much heavier, up to $0.3  $ g for anti-ice meteorite)
  antimeteorites might be suppressed and maybe almost absent in solar system; previous
bound by annihilation on the Earth may be considered for heavier
(10-100 milligram or above) anti-meteorites leading to a ratio
($r= 10^{-8}$) of antimatter allowable. Bounds by microgram
anti-meteorite annihilation on Moon soil while being very hard
and sharp, will be no more effective than the terrestrial bounds.
Moreover, there are other processes that may dilute above
antimeteorite presence in our solar system.

\section{THE ANTIMETEORITE ANNIHILATION and DECELERATION IN GAS}

Antimeteorite with a mass heavier than milligram may survive
annihilation: however while crossing a gas cloud, their lateral
annihilation may heat a meteorite side, leading to a rocket
ejection able to decelerate and at large matter gas density
gradient even divert and bounce the trajectory. However, for realistic density gradients the latter case can not
be realized and the momentum transfer due to annihilation
causes the antimeteorite deceleration in matter gas, which
can be described as follows.
Antimetorite of radius $r$, moving with a velocity $v$ in the
central field of gas, distributed around the central mass $M$
isotropically as
\begin{displaymath}
 \rho = \rho_{0} \cdot \ffrac {R_{0}}{R} ^{2}
\end{displaymath}
experiences the friction force due to annihilation

\begin{displaymath}
 F_{f} = - \rho(R) \pi r^{2} \eta v c
\end{displaymath}

where $\eta$ is the effectiveness of momentum transfer near unity;
assuming an initial anti-meteorite velocity $v_{a_i}$ and density
$ \rho_{a}$ and a normal galactic-disk mass density $ \rho $ one
finds the characteristic relaxation time $\tau$ (for a millimeter
anti-meteorite radius) :

\begin{displaymath}
          \tau = \frac{4}{3} \frac{\rho_{a}}{\rho}\frac{r}{\eta c}
          \end{displaymath}
\begin{equation}
 = 1.3 \eta^{-1}\cdot10^{2}\cdot year \frac{r}{mm} \frac{\rho_{a}}{g cm^{-3}} \frac{10^{-24} g cm^{-3}}{\rho}
\end{equation}
 Therefore in a short (in galactic scales) times any fast anti-meteorite will be slow
 down to a velocity comparable with common galactic gas. Therefore lightest
 anti-meteorite will follow a co-moving pattern with matter in
 galactic disk. Heavier ones (m $>> 0.1 $ g) will not evaporate and might reach the Earth.
 In presence of any radial gravitational force, near stars or star clusters, the gravitational force
\begin{displaymath}
 F_{g} = \frac{4}{3} \cdot \frac{GM \pi \rho_{a} r^{3}}{R_{0}^{2}}
\end{displaymath}
 and the friction action leads to a slow-down free fall up  to a steady value.
The equality of the two forces indeed leads to the constant
velocity
\begin{equation}
v = \frac{2}{3 \eta} \frac{\rho_{a}}{\rho_{0}} \frac{r}{R_{0}} \frac{R_{g}}{R_{0}} c
\end{equation}
 where
\begin{displaymath}
R_{g} = \frac{2GM}{c^{2}}
\end{displaymath}
is the Schwarchild radius of any central body.

The annihilation friction is effective, resulting in the
anti-meteorite deceleration and successive slow drift and final
annihilation towards the star center.

 In nearly horizontal motions the fast anti-meteorite may bounce
on the star-planet atmosphere and they may escape from the
central field. In the case of general motion and matter gas
distribution this effect may be estimated by assuming that a
fraction of antimatter is annihilated leading to a momentum
exchange (See \cite{Fargion})
and a velocity loss
 $\Delta v \sim v \sim 10^{-3} c$:

\begin{displaymath}
 \Delta v = \eta\cdot E/Mc
\end{displaymath}

 where $\eta$ is the fraction of annihilation energy going into
 effective anti-asteroid momentum exchange.
 Being necessary to escape from the galactic plane or from solar atmosphere
 a $\Delta v >10^{-3} c$ one finds

\begin{displaymath}
 (\Delta E)/(M c^2) = (\Delta M)/M \leq 10^{-3}/ \eta
\end{displaymath}

 This value cannot exceed unity otherwise the anti-meteorite will be
 totally annihilated; therefore the $\eta$ efficiency cannot
be below $10^{-3}$ but its value is bounded by the ratio of the
interaction length of charged pions on the meteorite volume; the
300 MeV pion crosses nearly 85 cm in water before interacting;
the total amount of matter crossed during meteorite life-time
traveling (comparable to galactic age) in the galactic disk is
nearly $10^{-2}$ g or $10^{-2}$ cm. of water. However in the case
of atomic antinuclei composition annihilating with hydrogen of
galactic gas the main consequence will be a breakdown of
antinuclei. Its fragments will deposit in a very efficient way
(nearly 50\%) the energy of annihilation into linear momentum as
well as increasing the temperature of the solid antimatter body.
Our first estimation show that the effective cooling is keeping
the temperature below the solid (rock) melting point, while the
antimeteorite moves in the Galaxy and Solar System. The
equilibrium temperature is established, provided that the heating
rate $2 \pi r^{2} \kappa \rho c^{2} v$ (where $\kappa$ is the
fraction of the total energy, released in the annihilation
($E_{an} = 2 m_{H} c^{2}$)), that heats the spherically symmetric
antimeteorite of radius $r$, moving with velocity $v$ in the
matter gas of density $\rho=m_{H} n$) is equal to the rate of
radiative cooling $4 \pi r^{2} \sigma T^{4} c$ (where $\sigma$ is
the Stephan-Boltzmann constant). In the considered approximation
both heating and cooling are proportional to the surface area, so
that the equilibrium temperature is given by $T_{e} = 168 K (n
\kappa v)^{1/4}$ for matter gas number density $n=$ $1 cm^{-3}$
and anti-meteorite velocity  v $= 300$ km/s. Annihilation of
matter gas with antinuclei on the antimeteorite surface leads to
its erosion, but its effect, which may deserve special analysis
for particular antimeteorite composition, does not lead to
significant change of the above estimation for sufficiently large
antimeteorites. Nevertheless the "ice" anti-comets might be melt
efficiently still in the galaxy and very efficiently near Solar
and Terrestial atmosphere. The reason is that the estimated value
of $T_{e}$ can easily be factor of 2 larger, but the
antimeteorite, moving with the velocity $v/c \approx 10^{-3}$,
with the account for all the uncertainties can be hardly heated
up to 1000 K due to the annihilation in the low density matter
gas (with the number density $n \approx 1 cm^{-3}$. The equilibium
condition, rewritten for energy density of radiation
($\epsilon_{\gamma} = 2.7 T n_{\gamma}$) and of annihilation
products ($\epsilon_{an} = 2 n m_{H} c^{2}$) in the form $
\epsilon_{\gamma} c \approx \kappa \epsilon_{an} v$, is reached
at $T_{e} \le 300$K due to the low values of in-flow velocity
$v/c \approx 10^{-3}$ and matter gas density $n/n_{\gamma}
\approx 10^{-9}$, what compensates the large value of
annihilation energy release $\frac{2 m_{H} c^{2}}{T_{e}} \le
2\cdot 10^{10}$.

\section{ANNIHILATION OF ANTI-ASTEROIDS on SUN}

The "galactic anti-asteroid" rate on Sun from (1) is
\begin{equation}
  \frac{dN}{dt} = 10^{10} r \ffrac{g}{M} year^{-1}
\end{equation}
The consequent event rate for suppressed anti- asteroids one over
a billion is 10 events a year. The fluence  $F$ on Earth is $3 \cdot
10^{-7} erg/cm^{2}$ and comparable to  GRB fluence, with a time
dilution of nearly 10 seconds. Therefore it may be well be missed
or misunderstood as a low energy solar flare. The rarest events at
100 g range may mimic observed solar flares. Let us remind that
present bounds in solar flare activity may be even detectable at a
nano-flare intensity. If the above coincidence is not just the
hint of the antimatter meteorites in-fall it provides the present
most stringent bound on antimatter. It may be useful to mention
that the two anti-meteorite searches undertaken in USSR in late
1960-s early 1970-s, even with no confirmation, exhibited the
positive effect ([13], see review in [14]) \footnote{see also in:
B.Konstantinov's Memorial Collection of works, Leningrad.
Fiz.Tech. Publ. (1974)}. So not only stringent limits, but even
positive discoveries should be in principle  considered in future
of such searches.
\section{CONCLUSIONS}
Anti-meteorites annihilations may provide the challenge to search
for antimatter in our Galaxy at the same level of sensitivity
which is planned to be reached in AMS-II experiment (a part over
a billion). With all the uncertainty in possible relationship
between the total mass of antimatter stars and the expected
amount of pieces of antimatter to be ejected by antimatter
stellar systems and all the possible reservations our first
estimate on Earth and Solar events are showing rather high
sensitivity ($10^{-8}-10^{-9}$) in antimatter search can or even
might be already  reached.

\section{ACKNOWLEDGEMENTS}
The work was partially performed in the framework of Russian State contract $40.022.1.1.1106$, with partial support of
Cosmion-ETHZ, AMS-Epcos collaborations, of grant $00-15-96699$
of support for Russian scientific schools, RFBR grant $02-02-17490$ and of grant of Russian Universities. One of us (M.Yu.Kh)
expresses his gratitude to Rome University La Sapienza and to IHES
for hospitality.


\begin{thebibliography}{10}
\bibitem{Cohen98} A.G. Cohen, A.De Rujula and S.L. Glashow,
 Astrophys. J., 495, 539 (1998)
\bibitem{Steigman76} G. Steigman, Rev. Astron. Astrophys., 14, 339 (1976)
\bibitem{Khlopov87} M.Yu. Khlopov, V.M.Chechetkin, Sov.J. Part.Nucl., 18, 267 (1987)
\bibitem{Sakharov67} A.D. Sakharov, JETP Lett., 5, 24 (1967)
\bibitem{Khlopov99} M.Yu. Khlopov, Cosmoparticle physics, World Scientific (1999)
\bibitem{Chechetkin82} V.M. Chechetkin , M.Yu. Khlopov, M.G. Sapozhnikov, Ya.B.
Zeldovich, Phys. Lett., 118 B, 329 (1982)
\bibitem{Khlopov00} M.Yu. Khlopov, S.G. Rubin, A.S. Sakharov, Phys.Rev., D63,
083505 1 (2000).
\bibitem{Khlopov00b} M.Yu. Khlopov, R.V.Konoplich, R.Mignani, S.G. Rubin , A.S.Sakharov,
Astroparticle Physics, 12, 367 (2000).
\bibitem{Khlopov00c} M.Yu. Khlopov, Grav.~\&~Cosm., 4, 69 (1998)
\bibitem{Golubkov00} Yu.A. Golubkov, M.Yu. Khlopov, Grav.~\&~Cosm., Supplement, 6, 104,(2000).
\bibitem{Fargion2001} D.Fargion, astro-ph/0002453, v570 n2 ApJ May
 10, 2002; 27th ICRC 2001, HE2.5,p.1297-1300; HE1.8,Germany,903-906,2001(2001)
\bibitem{Fargion} D.Fargion astro-ph/9803269
\bibitem{Konstantinov65} B.P.Konstantinov, G.E.Kocharov,
Doklady Akad. Nauk SSSR, 165, 63  (1965),
\bibitem{Chech82} V.M. Chechetkin , M.Yu. Khlopov, M.G. Sapozhnikov,
Rivista Nuovo Cimento, 5, N-10 (1982)
\end{thebibliography}
\end{document}